\documentclass[10pt,journal,compsoc]{IEEEtran}
\ifCLASSOPTIONcompsoc
  \usepackage[nocompress]{cite}
\else
  \usepackage{cite}
\fi
\ifCLASSINFOpdf
\else
\fi
\IEEEoverridecommandlockouts
\usepackage{amsmath,amssymb,amsfonts}
\usepackage{algorithmic}
\usepackage{graphicx}
\usepackage{textcomp}

\newtheorem{defn}{\textbf{Definition}}
\newtheorem{ruln}{\textbf{Rule}}

\begin{document}
\title{Microchain: A Hybrid Consensus Mechanism for Lightweight Distributed Ledger for IoT}

\author{Ronghua Xu,
        Yu Chen,~\IEEEmembership{~IEEE Senior Member},
        Erik Blasch,~\IEEEmembership{~IEEE Fellow},
        \\and Genshe Chen,~\IEEEmembership{~IEEE Senior Member},% <-this % stops a space
\IEEEcompsocitemizethanks{\IEEEcompsocthanksitem R. Xu and Y. Chen are with the Department of Electrical and Computer Engineering, Binghamton University, SUNY, Binghamton, NY, 13902, E-mail: \{rxu22, ychen\}@binghamotn.edu.
\IEEEcompsocthanksitem E. Blasch is with the U.S. Air Force Research Lab (AFRL), Rome, NY 13441, Email: erik.blasch@gmail.com.
\IEEEcompsocthanksitem G. Chen is with Intelligent Fusion Technology, Inc. Germantown, MD 20876, Email gchen@intfusiontech.com.}% <-this % stops an unwanted space
}

\IEEEtitleabstractindextext{%
\begin{abstract}
A blockchain and smart contract enabled security mechanism for IoT applications has been reported recently for urban, financial, and network services. However, due to the power-intensive and a low-throughput consensus mechanism in existing blockchain, like Bitcoin and Ethereum, there are still challenges in integrating blockchain technology into resource-constrained IoT platforms. In this paper, Microchain, based on a hybrid Proof-of-Credit (PoC)-Voting-based Chain Finality (VCF) consensus protocol, is proposed to provide a secure, scalable and lightweight distributed ledger for IoT systems. By using a bias-resistant randomness protocol and a cryptographic sortition algorithm, a random subset of nodes are selected as a final committee to perform the consensus protocol. The hybrid consensus mechanism relies on PoC, a pure Proof of stake (PoS) protocol, to determine whether or not a participant is qualified to propose a block, given a fair initial distribution of the credit assignment. The voting-based chain finality protocol is responsible for finalizing a history of blocks by resolving conflicting checkpoint and selecting a unique chain. A proof-of-conception prototype is implemented and tested on a physical network environment. The experimental results verify that the Micorchain is able to offer a partially decentralized, scalable and lightweight distributed ledger protocol for IoT applications.
\end{abstract}

}

\maketitle

\IEEEdisplaynontitleabstractindextext

% \fi
\IEEEpeerreviewmaketitle

%======================================== introduction =======================================
%\IEEEraisesectionheading{\section{Introduction}\label{sec:introduction}}
\section{Introduction}
\label{sec:intro}  % \label{} allows reference to this section
The prevalence of Internet of Things (IoT) allows heterogeneous and lightweight smart devices to collaboratively provide services with or without human intervention. With an ever-increasing presence of IoT-based smart applications and their ubiquitous visibility from the Internet, the highly connected smart IoT devices with a huge volume of generated transaction data incur more concerns on security and privacy \cite{chen2018smart}. IoT systems are deployed in a distributed network environment that includes a large number of IoT devices with high heterogeneity and dynamics. The heterogeneity and resource constraint at the edge necessitate a scalable, flexible and lightweight system architecture that supports fast development and easy deployment among multiple application vendors using non-standard development technology. Furthermore, those smart devices are geographically scattered across near-site network edges, and managed by fragmented service providers that enforce different security policies. Thus, traditional security policies on a centralized authority basis, which suffer from the performance bottlenecks or single point of failures, are not efficient and suitable to address performance and security challenges in IoT systems.

Recently, designing new decentralized security mechanisms for distributed network applications becomes one of the most intensively studied topics in academics and industries. Blockchain, which acts as the fundamental protocol of Bitcoin  \cite{nakamoto2008bitcoin}, has demonstrated great potential to revolutionize the fundamentals of information technology (IT) due to many attractive properties, such as decentralization and transparency \cite{novo2018blockchain}. The blockchain enabled security mechanisms for IoT-based applications have been reported recently, for example, smart surveillance system \cite{nagothu2018microservice, nikouei2018real}, social security system \cite{xu2018constructing}, space situation awareness \cite{xu2018exploration}, biometric imaging data processing \cite{xu2019decentralized}, identification authentication \cite{hammi2018bubbles} and access control \cite{xu2018blendcac, xu2018smartcac}. Blockchain and smart contract together are promising to provide a decentralized security mechanism to IoT systems.

Integrating blockchain technologies into IoT systems presents critical challenges in designing scalable and lightweight blockchain protocols. Particularly, the performance of blockchain networks significantly relies on the performance of consensus mechanisms, e.g., in terms of data consistency, speed of consensus finality, robustness to arbitrarily behaving nodes, and network scalability \cite{wang2019survey}. Unfortunately, existing blockchain protocols are not suitable to be directly merged to IoT scenarios. Most of the existing permissionless blockchain networks (e.g., Bitcoin) require solving a time-consuming hashing puzzle for block generation. Such hashing-intensive proof-of-work (PoW) consensus mechanisms could guarantee the network scalability and mitigate Sybil attacks; however, they are at the cost of low throughput, high energy consumption and ever-increased chain-data. The classical Byzantine consensus protocols, such as Practical Byzantine Fault Tolerance (PBFT) \cite{castro1999practical}, demonstrate a better performance with high throughout, lower latency and limited overhead. However, they rely on identity authentication and allow very limited network scalability in terms of the number of nodes. 

To address challenges in integrating blockchain technologies into IoT systems, we propose Microchain, a hybrid Proof-of-Credit (PoC)-Voting-based Chain Finality (VCF) blockchain architecture that provides a secure, scalable and lightweight distributed ledger. Following the idea of delegation, microchain chooses a small subset of the nodes in the network as validators, and those validators form a final committee, called dynasty, to perform a consensus function. The verifiable random function combining with cryptographic sortition provide a bias-resistant randomness foundation to periodically form a statistical representative dynasty. The proposed novel PoC-VCF based final committee consensus mechanism not only offers higher efficiency over protocols based on proof of physical resources like PoW, but it is also more scalable than those classical BFT enabled consensus solutions. In summary, this paper makes the following contributions:

\begin{itemize}
\item[1)] A systematic architecture of a microchain is introduced, and key components and work flows are explained;

\item[2)] A novel PoC-based block proposal mechanism is designed to improve energy efficiency and throughput of PoW-based blockchain;

\item[3)] A lightweight voting-based chain finality protocol is developed for resolving the chain fork problem in chain extension; and 

\item[4)] A proof-of-concept prototype is implemented as a private blockchain network and tested on Raspberry Pi, and the experimental results validate the feasibility of running microchain on IoT systems with lower energy assumption and communication overhead. 
\end{itemize}
%The PBTF-based intra-committee consensus algorithm allow validators of within the shard to commit data transactions quickly, thus, improve performance with higher throughput and lower communication overhead. The Proof of Credit (PoC), which is a variant of PoS, is performed among members in final committee to finalize global state and extend blockchain.  

The remainder of this paper is organized as follows: Section \ref{sec:relatedwork} reviews background knowledge of blockchain consensus protocols. Section \ref{sec:architecture} introduces the microchain architecture and describes the core features. A novel PoC-based block proposal mechanism is explained in Section \ref{sec:PoC} and Section \ref{sec:BFT} illustrates the voting-based chain finality protocol. The experimental results are discussed in Section \ref{sec:experiment}. Finally, a summary is presented in Section \ref{sec:conclusion}. 

%======================================== related works =======================================
\section{Background and Related Work}
\label{sec:relatedwork}  

As one of the most fundamental problems in fault-tolerant distributed computing, \emph{consensus} states that the processes have to reach agreement in a non-trivial way. This agreement takes the form of a value (called decision value) that has to be one of the values proposed by the processes to the consensus instance \cite{raynal2010communication}. In the context of a distributed computing environment, the consensus in blockchain networks can be defined as a fault-tolerant state-machine replication problem by maintaining the global blockchain state across the P2P network. Each participating node maintains a local replicate of the blockchain and shares it with other nodes. All nodes make an agreement (consensus) on the unique common replicate of the blockchain in the condition of Byzantine failures \cite{schneider1990implementing, lamport1982byzantine}. Given variant consensus protocols, the blockchain networks could be categorized into permissionless blockchain (e.g., Nakamoto Consensus Protocols) and permissioned blockchain (e.g., Practical Byzantine Fault Tolerant Consensus).   

\subsection{Nakamoto Consensus Protocols}

To jointly address the critical issues, such as pseudonymity, scalability and poor synchronization in an open-access network environment, the Nakamoto consensus protocol \cite{nakamoto2008bitcoin} is based on a Proof of Work (PoW) scheme. It is implemented as the consensus foundation of Bitcoin, and widely adopted by other cryptocurrency-based blockchain networks. PoW is essentially an incentive-based consensus protocol in which participants compete for rewards through a cryptographic block-discovery racing game. Nakamoto protocol introduces incentives to probabilistically award the consensus participants to ensure proper functioning of a permissionless blockchain network \cite{nakamoto2008bitcoin}. The Nakamoto consensus protocols demonstrate scalability in a trustless open-access network environment. However, PoW also incurs several issues, such as limited throughput, high demand of computation and storage resources as well as unsustainable energy consumption. 

To reduce energy consumption, Proof-of-Stake (PoS) was proposed by Peercoin \cite{king2012ppcoin}. Within a Peercoin network, a metric of ``coin age'' is proposed to measure a miner's stake by multiplying the held tokens and the holding time. The PoS kernel protocol allows a miner to use its stake to solve the puzzle solution, the probability of proposing a new block follows the stake distribution. Compared to PoW protocols that use brute-force hashing power to solve computation-intensive puzzle solutions, PoS leverages the distribution of token ownership to simulate a verifiable random function to propose new blocks. Since the block miners only consume limited resources, PoS is also known as a process of ``virtual mining'' \cite{bonneau2015sok}.

%Based on the framework of Nakamoto protocol, a 
A number of alternative Proof of Concept (PoC) schemes have been proposed to improve performance of existing PoW in terms of security, fairness and sustainability. To address the issues of centralized computation power pool, the Proof of Memory (PoM), a memory-hard PoW is adopted by ZCash \cite{hopwood2016zcash} and Ethereum \cite{buterin2014next} networks. With the purpose of useful resource provision, the idea of ``Proof of Useful Resources'' (PoUS) has been proposed to tackle the resource wasting problem of PoW \cite{wang2019survey}. The Proof of Exercise (PoE) is proposed to replace the computation intensive searching problem in PoW with the useful ``exercise'' of matrix product problems \cite{shoker2017sustainable}. Based on Trusted Execution Environment (TEE), Resource-Efficient Mining (REM) \cite{zhang2017rem} verifies and measures the software running in an Intel Software Guard Extensions (SGX)-protected enclave that randomly determines whether the work leads to a valid block proof. 
%Apart from proof of useful computation, Permacoin \cite{miller2014permacoin} proposes an alternative scratch-off puzzle for Bitcoin based on Proofs-of-Retrievability (POR) in term of distributed storage provision. The consensus scheme requires random access to a subset of the locally stored segments and and the corresponding Merkle proof. Given the competition among mining clients in Bitcoin, the Permacoin based on POR gives rise to highly decentralized file storage, thus reducing the overall waste of Bitcoin \cite{miller2014permacoin}.

\subsection{Byzantine Fault Tolerant Consensus}

The consensus in a distributed system can be expressed abstractly as a Byzantine General Problem \cite{lamport1982byzantine}, which copes with single value agreement among different parts of a system given failure of communication or conflicting information. The problem is solvable if and only if more than two-thirds of the generals are loyal under oral messages. Thus, a consensus protocol is called Byzantine Fault Tolerant (BFT) if it can tolerate one or more byzantine failures while making a correct agreement. For any BFT consensus protocol, the distributed network configuration should satisfy the following requirement: $N\geq3f+1$, where $f$ is the number of Byzantine processes. The PBFT algorithm for Statement Machine Replication (SMR) \cite{lamport1978time} that tolerates Byzantine faults \cite{castro1999practical} and has been widely adopted as a basic consensus solution in the blockchain community, like Hyperledge Fabric \cite{hyperledgerfabric}. The PBFT algorithm offers both liveness and safety in asynchronous network environment given assumption that at most of $\lfloor\frac{n-1}{3} \rfloor$ out of total of $n$ replicas are Byzantine faults. All peering nodes in PBFT are categorized as either a leader node or some validating nodes. These peers will execute three phases of consensus protocol, called Pre-prepare, Prepare and Commit, to make agreement on proposed new blocks. As long as more than 2/3 of all nodes are honest and they have successfully executed the Commit phase, the transactions recorded in the validated blocks are appended to the current chain.

To tackle the weakness of consensus algorithms that suffer from high latency due to the requirement that all nodes within the network communicate synchronously, Ripple \cite{schwartz2014ripple} is proposed to circumvent this requirement by utilizing collectively-trusted subnetworks within the larger network. 
In Ripple, all verified transactions will be added directly to the raw ledger rather than through the chain of blocks. To make the verification, each node maintains their own peer list, called a unique node list (UNL), to broadcast the transaction to such subnetworks instead of to nodes of whole network. The intersection of UNL by any two different peers should be at least 1/5 of all the nodes over the network. Since the ``trust'' required by these subnetworks is in fact minimal and can be further reduced with principled choice of the member nodes. 
The Ripple provides a low-latency consensus algorithm which still maintains robustness in the face of Byzantine failures \cite{schwartz2014ripple}. As a variant of Ripple, Steller \cite{mazieres2015stellar} designed a new model for consensus called federated Byzantine agreement (FBA), which ensures nodes agree on slot contents. 
Similar to UNL in Ripple, Steller utilizes quorum slices for consensus agreement made between trust nodes, which is the same as validators in Ripple. In Steller consensus process, sub-network level trust decisions are made by nodes in a quorum, which is a set of nodes sufficient to reach agreement. Then, the subset of a quorum work together as a federated committee to determine system-level agreement. 
Compared to decentralized proof of-work and proof-of-stake schemes, Steller has modest computing and financial requirements, lowering the barrier to entry and potentially opening up financial systems to new participants \cite{mazieres2015stellar}.

\subsection{Hybird Consensus Protocol}

To improve limited performance of permissionless consensus without undermining the unique features such as scalability, combining a scalable permissionless consensus (eg., PoW) with a high throughput permissioned consensus (e.g, BFT) becomes a prospective approach. The Bitcoin-NG \cite{eyal2016bitcoin} is proposed to improve performance of PoW-based Nakamoto protocol by by decoupling Bitcoin’s blockchain operation into two planes: leader election and transaction serialization. The protocol divides time into sequential epochs, and only a single leader is in charge of serializing state machine transitions in each epoch. To bootstrap the transaction throughput, the protocol introduces two types of blocks, namely, the key blocks that require a PoW puzzle solution for leader election and the microblocks that require no puzzle solution and are used for transaction serialization \cite{wang2019survey}. Although Bitcoin-NG may also experience key block forks, it scales optimally with bandwidth limited only by the capacity of the individual nodes and latency limited only by the propagation time of the network \cite{eyal2016bitcoin}.

To design a computationally-scalable Byzantine consensus protocol for blockchain, SCP \cite{luu2015scp} is proposed through incorporating BFT and sharding into blockchain consensus. The key ideas are inspired by concept of "sharding" \cite{croman2016scaling} in infrastructures of distributed database and cloud. Through securely establishing identities for network participants, whole network are randomly divided into several sub-committees. Each sub-committee performs a classical BFT consensus protocol to process a separate set of transactions and propose blocks in parallel. A final committee is designated to combine the outputs of sub-committees into an ordered blockchain data structure. To extend existing consensus protocol based on SCP, Elastico \cite{luu2016secure} is proposed to build a secure sharding protocol for open blockchains. In a epoch, the candidates of committees attempt to find a PoW puzzle solution provided a seed called "epochRandomness" that is a public random number string generated in previous epoch. ELASTICO exhibits almost linear scalability throughput with computation capacity with roughly $O(n)$ message complexity. However, the participants have to download full blockchain data to perform the consensus task, which brings latency in bootstrapping process and storage overload on client nodes.

To enable parallelization of both network consensus and data storage, a "full sharding" protocol called named
"OmniLedger" \cite{kokoris2018omniledger} is designed to provide "statistically representative" shards for permissionless transaction processing. OmniLedger uses a bias-resistant protocol called RandHound \cite{syta2017scalable} to generate epoch global randomness strings for sharding committees formation. To optimize trade-off between the number of shards, throughput and latency, the intra-shard consensus follows a "Optional Trust-but-Verify Validation" model, where optimistic validators make a provisional but unlikely-to-change commitment and core validators subsequently verify again the transactions to provide finality and ensure verifiability \cite{kokoris2018omniledger}. To secure corss-shard transactions, OmniLedger introduce a novel Byzantine Shard Atomic Commit protocol to handle atomically transactions processing acorss shards. Furthermore, a gradually in-and-out committee members swap strategy could reduce extra message overhead and bootstrapping latency in shard reconstruction. Another epoch-based, two-level-BFT protocol called RapidChain \cite{zamani2018rapidchain} is proposed for scaling blockchain vie full sharding.  RapidChain employs block pipelining strategy to achieve very high throughputs in intra-committee consensus. Further more, a novel gossiping protocol for large blocks reduces large overhead on committee-to-committee communication, and ensure an efficient cross-shard transaction verification.

%======================================== system design =================================
\section{System Design of Microchain}
\label{sec:architecture}  
The goal of the Microchain is to enable a lightweight distributed ledger system in resource constrained IoT environments at the network edge, which is achieved by introducing an efficient consensus mechanism running on a small number of validators. The microchain network proceeds transactions by a final committee in a fixed time periods called dynasty epoch. A random committee formation protocol ensures that committee selection process is unpredictable. In each dynasty life time, a hybrid PoC-VCF consensus mechanism is responsible for proposing block and finalizing chain history given a unbounded time delay.
The final-committee consensus which is built on a proof-of-stake based finality mechanism. A Proof-of-Credit (PoC) protocol, which is a pure PoS mechanism, determines whether a participant is selected to propose a block given fair initial distribution of the credit stake to the committee members in a given epoch. While a Voting based chain finality (VCF) mechanism could protect against fork by resolving conflicting checkpoints and finalize the history of blockchain. 

The microchain assumes a synchronous network in which operations of processes are coordinated in rounds with bounded delay constraints, and provides two formal proprieties of a robust distributed ledger: 
\begin{itemize}
\item \emph{Persistence}: ensures a safety goal that all users should agree on the same transactions and finalized transactions should be in the same position in the distributed ledger. If a honest node of the system accepts a transaction $tx$ as finalized in block $B_i$, then any other nodes could query $tx$ in block $B_i$.
\item \emph{Liveness}: ensures that transactions submitted by honest nodes are confirmed in finalized blocks after a sufficient amount of time.
\end{itemize}
 %This work mainly focus on final-committee consensus mechanism.
 
\subsection{Prerequisites}
Essentially, a distributed ledger is a partial ordering of the block generating events, thus blocks followed the partial relation denoted by '$\rightarrow$'. For two blocks $a$ and $b$, if $a$ comes before $b$, then $a \rightarrow b$. Given assumption of synchronous network environment, the time is dividend into discrete units called $slots$ to measure upper bounded delay in block proposal and finality operations. In a distributed system, it is hard to say that one of two events occurred first given synchronous time discrepancies among participants. Therefore, we introduce logical clocks called $ticks$, which can be used to totally order the events \cite{lamport1978time}. In general, each slot $sl_t$ is indexed by tick $t$, where tick could be represented as monotonically increasing integers $t \in \{1,2,3, ...\}$. To ensure liveness, the length of time window represented by a slot should be sufficient to guarantee that message transmitted by a sender will be received by its intended recipient within that time window (accounting for local time discrepancies and network delays). The upper bounded delay is defined as $\Delta$, then $sl_t \geq \Delta$, where $t \in \{1,2,3, ...\}$. Each participant must finish registration process to join the permissioned blockchain network. A asymmetric key pair generation functionality creates a signing-verification keys pair $(sk_i, pk_i)$ for user $u_i$, and user's credit stake $c_i \leq C_{max}$, where $C_{max}$ is maximum value of credit stake, is associated with user's public key $pk_i$. All registered users in system can be represented as a set $U=\{(pk_1,c_1), (pk_2,c_2),..., (pk_n,c_n) \}$, where $n$ is the registered user's number.

\subsection{Basic Concepts}
Before introducing the microchain architecture, several basic definitions are defined as following:

\begin{defn}
Epoch - An epoch is a set of $R$ sub-sequential slots, represented as $e=\{sl_1, sl2,...,sl_t\}$, where $0 \leq t \leq R$. The $R$ value is epoch size that is multiple unit slot $sl$.
\end{defn}

\begin{defn}
Validator - A validator $v_i \in V$ ($V \subseteq U$) is a registered user who is qualified for final-committee selection and performing a consensus algorithm.
\end{defn}

\begin{defn}
Dynasty - A dynasty is a set of current final-committee members $D=\{(pk_1,c_1), (pk_2,c_2),..., (pk_k,c_k) \subseteq V \}$, where $0 \leq k \leq K$, and the $K$ value is committee size. The dynasty is selected from validator set $V$ based on final committee selection algorithm, and dynamic life time is multiple epochs.
\end{defn}

\begin{defn}
Block - A block generated at a slot $sl_t$ ($t \in {1,2,3,..}$) by validator $v_j$ is represented as $B_i=( pre\_hash, height, tx\_data, sl_t, pk_j, \sigma_j)$. where the parameters are:
\begin{itemize}
\item $pre\_hash$: is a $\lambda$-bit-length hash string of previous Block $B_{i-1}$, which is represented as $\mathcal{H}(B_{i-1})$=$\mathcal{H}(pre\_hash, height, tx\_data, sl_{t-1})$, where $\mathcal{H}(\cdot)$ is a pre-defined collision-resistant hash function that output hash string $h \in \{0,1\}^{\lambda}$;
\item $height$: height of current block in blockchain (ledger);
\item $tx\_data$: transactions data $d \in \{0,1\}^*$;
\item $sl_t$: block generated time stamp at $sl_t$;
\item $pk_j$: public key of validator that proposes the block; and
\item $\sigma_j$: a signature $Sign_{sk_j}(tx\_data, pre\_hash, height, sl_t)$ signed by sender $pk_j$.
\end{itemize}
\end{defn}

\begin{defn}
Genesis Block - The genesis block is a special block which is defined as $B_0=(pre\_hash=0, height=0, sl_0=0, init\_D)$, where $init\_D$ is the initial dynasty.
\end{defn}

\begin{defn}
Blockchain - A blockchain (or public ledger) $\mathcal{C}$ is a partial order of blocks $B_0 \rightarrow B_1 \rightarrow ... \rightarrow B_{n-1} \rightarrow B_n$ associated with a strictly increasing sequence of slots $sl_t$. Each block $B_i$ use $pre\_hash$=$\mathcal{H}(B_{i-1})$ to chain previous block $B_{i-1}$. The significant parameters are described as follows:
\begin{itemize}
\item $length$: is the length of chain denoted $len(\mathcal{C})=n$ to count number of blocks between genesis block $B_0$ and confirmed $B_n$;
\item $head$: indicates the head of chain denoted $head(\mathcal{C})=B_n$, where $B_n$ is last confirmed block that is extended on finalized main chain.
\end{itemize}
\end{defn}

Given the above definitions, the next sub-section explains how the final committee consensus protocol works on a microchain architecture.

\subsection{Architecture  Overview}

%fig2_final_commitee_consensus
\begin{figure} [t]
\begin{center}
\begin{tabular}{c}
\includegraphics[height=6.5 cm]{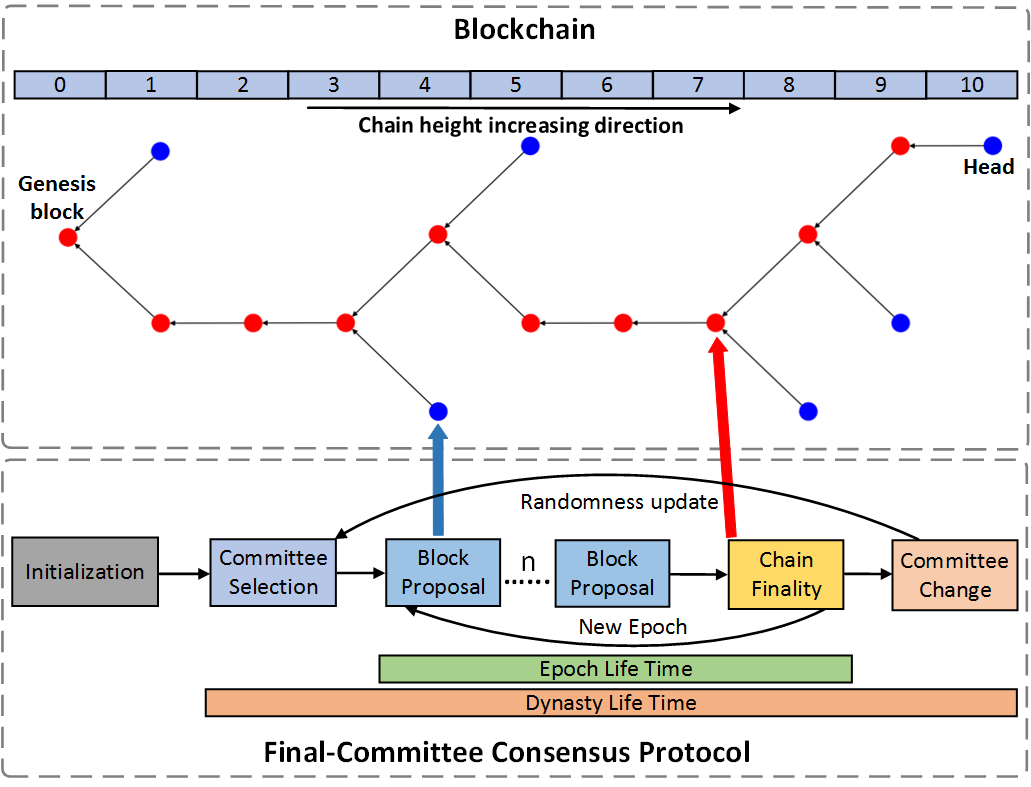}
\end{tabular}
\end{center}
\caption[example] { \label{fig:final_consensus} Microchain architecture overview.}
\vspace{-10pt}
\end{figure}

The blockchain structure is illustrated as upper Fig. \ref{fig:final_consensus}. The blue nodes represent confirmed blocks while red ones indicate finalized blocks. The genesis block is the root node of blockchain, and each block use its $pre\_hash$ to point to parent block and extend chain. The chain height follows a strictly increasing sequence of finalized blocks (path through red nodes). The head of blockchain is a last confirmed block whose parent is finalized and has largest height. The lower Fig. \ref{fig:final_consensus} demonstrates key components in microchain network, and the work flows are described as follows:

\begin{itemize}
\item \emph{Initialization}: In initialization process, a special dynasty, which includes a group of validators specified by administer, acts as initial committee $D_{init}$ to initialize blockchain. Each validator creates genesis block $B_0$ and sets the local blockchain $\mathcal{C}=B_0$ and $head=B_0$. The initial committee will work as the first dynasty of system until following dynasty change finished. Thus, $B_0$ is beginning block of initial dynasty epoch.

\item \emph{Committee Selection}: If the current dynasty is the initial committee, then skip this step. Otherwise, at the beginning of new dynasty life time, the final-committee formation protocol exploits a  Verifiable Random Function (VRF) based cryptographic sortition scheme \cite{gilad2017algorand} to randomly choose a subset of validators $V$ as final-committee according to their credit weight.
%a final committee selection based on cryptographic sortition algorithm is performed among validators set $V$, which includes all candidates of final-committee. 
The selected committee members $D$ will be added to current block, which is marked as beginning block of new dynasty epoch. The life time of dynasty epoch starts from committee selection and ends after dynasty change.

\item \emph{Block Proposal}: The block proposal mechanism uses a pure PoS protocol, called Proof-of-Credit (PoC), to generate new blocks in each block proposal run. Only validators in the current dynasty is able to propose new block. The probability that a user could proposes a block is associated with its credit distribution of current dynasty. If validator $v_j$ could solve puzzle difficulty problem in slot $sl_{t+1}$ by computing $\mathcal{H}(B_{i}, pk_j, c_j) \leq d_{cond}$ ($d_{cond}$ is difficulty condition target value), it generates new block $B_{i+1}=(\mathcal{H}(B_{i}), height+1, tx\_data, T_{stamp}, pk_j, \sigma_j)$ and broadcasts it with a valid signature to all committee members of the current dynasty. Each committee member accept all valid blocks in the current slot, and verifies if blocks meet confirmation requirements. The verified  block will be added to local chain $\mathcal{C}$ with $head=B_{i+1}$.

\item \emph{Chain Finality}: At the end of an epoch, the $head$ with epoch height becomes a checkpoint which is used to resolve forks and finalize chain history. The chain finality uses a voting-based algorithm to commit checkpoint block and finalizes those already committed blocks on the main chain. The chain finality ensures that only one path including finalized blocks becomes a main chain, as Fig. \ref{fig:final_consensus} shows. Therefore, the following blocks in new epoch are only extended on such unique main chain. The chain fork problem is prevent by resolving conflicting checkpoints and finalizing the history of the blockchain. 

\item \emph{Committee Change}: At the end of dynasty life time, the current committee members make agreement on a new dynasty randomness string. The epoch randomness string generation uses the RandShare mechanism to make agreement on proposing the next epoch randomness string among members of final committee. RandShare is an randomness protocol which is based on Publicly Verifiable Secret Sharing (PVSS) \cite{stadler1996publicly, schoenmakers1999simple} to ensure unbiasability, unpredictability, and availability in public randomness sharing. The proposed unbiasable and unpredictable public randomness string will be used for committee selection process of the next dynasty life time.
\end{itemize}

As two core functions in epoch life time of blockchain extension: block proposal and chain finality, detail designs are explained in following sections.

%---------------------------------------------- Block Proposal Mechanism --------------------------------------
\section{PoC-based Block Proposal Mechanism}
\label{sec:PoC}  
The PoC-based block proposal mechanism is mainly responsible for generating blocks and extending chains. Essentially, our PoC-based block proposal mechanism is a pure chain-based proof of stake, which mimics proof of work mechanics and simulates mining by pseudorandomly assigning block proposal right to validators. The block generation relies on a slot leader selection process which is associated with credit distribution $\mathcal{D}$ of current dynasty. A chain extension rule ensures that new blocks are growing on the finalized main chain and prevents against conflicting blocks.

\subsection{Transaction Pooling}

%Each user could send transactions to interact with blockchain network. A transaction is represented as $tx=\{tx\_hash, pk_{sender}, pk_{recipient}, T_{stamp}, data, \sigma \}$, where $tx \in \{0,1\}^*$. 
Each use could send transactions to interact with blockchain network. The transaction is defined as following:
\begin{defn}
(Transaction). A transaction is represented as $tx=\{tx\_hash, pk_{sender}, pk_{recipient}, T_{stamp}, data, \sigma \}$ ($tx \in \{0,1\}^*$), where:
\begin{itemize}
\item $tx\_hash$: is a $\lambda$-bit-length hash string of transaction $tx$, which is represented as $tx\_hash$=$\mathcal{H}(pk_{sender}, pk_{recipient}, T_{stamp}, data)$, where $\mathcal{H}(\cdot)$ is a pre-defined collision-resistant hash function that output hash string $h \in \{0,1\}^{\lambda}$;
\item $pk_{sender}$: is the sender's public key;
\item $pk_{recipient}$: is the recipient's public key;
\item $T_{stamp}$: is time stamp of generating transaction;
\item $data$: is information $d \in \{0,1\}^*$ enclosed by transaction;
\item $\sigma$: a signature $Sign(tx\_hash, pk_{sender}, pk_{recipient},$ $T_{stamp}, data)$ signed by sender's private key $sk_{sender}$.
\end{itemize}
\end{defn}

After receiving those transactions broadcasted on the network, each validator verifies received transactions according to the following conditions:
\begin{itemize}
\item[i)] sender and recipient are registered users in $U$, and $Verify(tx\_hash, pk_{sender}, pk_{recipient}, T_{stamp}, data) = \sigma$ by using sender's public key $pk_{sender}$;
\item[ii)] transaction $tx$ should neither be existed in transactions pool nor the last $\kappa$ committed blocks; and
\item[iii)] Time stamp $T_{stamp}$ must fall in either current time slot or past sequential slots of last $\kappa$ committed block;
\end{itemize}

The condition i) could prevent against transactions from invalid users or any malicious modification. While condition ii) and iii) are  mainly for preventing duplicated $tx$ in $\kappa+1$ period of time slots. After the verification process, only valid transactions are cached as local transactions pool denoted as $TX=\{tx_1, tx_2, ..., tx_N\}$, where $N$ is transactions pool size. The validator also uses condition iii) to regularly check local transaction pool to remove those outdated transactions, which has not been confirmed by the last $\kappa$ committed blocks.   

\subsection{Slot Leader Selection}
For each slot $sl_t$ in Epoch $e=\{sl_1, sl2,...,sl_R\}$, a random slot leader selection process determines if a validator $v_i$ is able to propose a new block for current block given the probability $p_i$ is proportional to its weight by credit $c_i$ in current dynasty. The credit distribution is defined as following:

\begin{defn}
Credit Distribution - is represented as $\mathcal{D}=\{ p_1, p_2,..., p_K\}$,  where $p_j= \frac{c_j}{\sum_{j=1}^{K}c_j} $ and $K$ is committee size. 
\end{defn}

To become a slot leader to propose new block, a validator must show its proof by solving a puzzle problem. Unlike PoW that utilizes a brute-force manner to find a nonce to meet the uniform target difficulty, each validator $v_j$ simply computes proof based on chain head block and its credit $c_j$. The target difficulty that is associated with its credit weight $p_j$ determines if proof is valid or not. The Proof-of-Credit (PoC) puzzle and solution can be formally defined as follows:

\begin{defn}
Proof-of-Credit - Given an adjustable difficulty condition parameter $\xi$, the process of PoC puzzle solution aims to verify a solution string $hc$, which is the hashcode of concatenation of $\mathcal{H}(head(\mathcal{C}))$, $pk_j$ and $c_j$, such that the value calculated by taking $\xi$ length lower bits of the $hc$ is smaller than a target value generated by difficulty condition $d_{cond}(\xi,p_j)$:

\begin{equation}
\label{eq:poc_puzzle}
\mathcal{TB}(hc, \xi) \leq d_{cond}(\xi,p_j)
\end{equation}
where $hc = \mathcal{H}(\mathcal{H}(head(\mathcal{C})),pk_j,c_j)$; $\mathcal{TB}(hc, \xi)$ function outputs lower $\xi$ bits of the hashcode $hc$; and difficulty condition function $d_{cond}(\cdot, \cdot)$ is denoted as:
\begin{equation}
\label{eq:target_condition}
d_{cond}(\xi,p_j) = (2^{\xi}-1) \cdot p_j
\end{equation}
where $d_{cond}(\xi,p_j) \in \{0,1\}^\xi$.
\end{defn}

Given the above definition, the PoC-enabled slot leader selection process is described as the following:

\begin{itemize}
\item \emph{Block Proposal}: during the current time slot $sl_t$, each validator $v_j$ in dynasty regularly calculates hashcode $hc$ based on current status of credit stake $c_j$, and use Eq. (\ref{eq:poc_puzzle}) to determine if it is qualified to propose a new block. If the candidate solution $\mathcal{TB}(hc, \xi)$ is smaller than target value defined by Eq.\ref{eq:target_condition}, then he generates a new block $B_{i+1}=(\mathcal{H}(head(\mathcal{C})), height+1, tx\_data, T_{stamp}, pk_j, \sigma_j)$ and broadcasts it to the network. The validator $v_j$ with higher credit weight $p_j$ is easier to calculate solution string $hc$ that falls into range of $[0,(2^{\xi}-1) \cdot p_j]$. Thus, he has the higher probability to become a slot leader to propose a new block. Furthermore, parameters used for calculating hashcode $hc$ are not changed in current slot time. Therefore, the $hc$ is a fixed value, and each validator is only allowed to propose one block in current slot time.

\item \emph{Block Verification}: after receiving proposed block $B_{i+1}$, each validator $v_j$ firstly checks if sender $pk_j$ is a valid committee member and verifies signature $\sigma$ by using the sender's public key $pk_j$. If block is from valid committee member with correct signature, then he will check if $T_{stamp}$ falls in time slot $sl_{t}$ duration, $B_{i+1}[height]=head(\mathcal{C})[height]+1$ and $B_{i+1}[pre\_hash]=\mathcal{H}(head(\mathcal{C}))$. Finally, validator verifies proof of solution to block $B_{i+1}$ based on the PoC algorithm defined in Eq.\ref{eq:poc_puzzle}. If all conditions are true, block is accepted as confirmed status and validator updates head of local chain as $head(\mathcal{C})=B_{i+1}$. Those sequential verification operations ensure that only valid blocks generated in the current time slot $sl_t$ with correct proof of solution are appended to head of local chain.
\end{itemize}

Compared to PoW-based leader selection process, like Bitcoin, our PoC have the following advantages:

\begin{itemize}
\item \emph{Energy efficiency}: Since input parameters for calculating $hc$ is fixed in one time slot, so a validator just executes the PoC mining task once to know if he is qualified for proposing new block in current leader selection round. Such mining method is more energy efficient than PoW-based solutions, which uses a computation-intensive mechanism to win the puzzle-solving race;

\item \emph{High Throughput}: The block proposal operation is only performed by validators in the current committee $D$ with limited  $K$ members, it's easier to ensure a synchronous network assumption that time slot $sl_t$ for block proposal rounds could be no more than upper bounded delay constraints $\Delta$. Given assumption that each block only encloses fixed amount of transactions with same data size, the block confirmation time is only related to length of time slot $sl_t$. The high throughput rate is achieved by correctly choosing a short $sl_t \geq \Delta$.
\end{itemize}

\subsection{Chain Extension Rule}

In general, the chain extension follows a "largest height of confirmed block" rule, which requires that new blocks $B_{i+1}$ are only appended to $head(\mathcal{C})$ by setting $B_{i+1}[pre\_hash]=\mathcal{H}(head(\mathcal{C}))$. Although slot leader selection process allows that the probability of block generated by validator $v_j$ is proportional to its credit weight $p_j$, it cannot ensure that only one block is proposed in the current block proposal time slot. The amount of candidate blocks $b$ could be between 0 and dynasty size $K$. Given candidate block number $b \in [0, K]$ in current time slot, the chain extension rules are described as follows:

\begin{itemize}
\item[i)]$b=1$: if there is only one proposed candidate block $B_{i+1}$, then block is accepted as confirmed status and update the chain head as $head(\mathcal{C})=B_{i+1}$;
\item[ii)]$b>1$: if more than one candidate blocks are proposed, then all blocks are accepted as confirmed status. The $head(\mathcal{C})$ update follows two sub rules: 1) chain head points to a block whose sender has the highest credit $c$ than other blocks' senders; and 2) if there more than one candidate blocks who have the same highest credit $c$, the block who has the smallest PoC condition value $\mathcal{TB}(hc, \xi)$ becomes the chain head; and
\item[iii)]$b=0$: if no proposed block is received before the end of current time slot, an empty block denoted as $B_{empty}=(pre\_hash, height+1, null, sl_t, null, null)$ is generated as $B_{i+1}$ and accepted as confirmed status. The chain head update process follows the rule i) $b=1$.
\end{itemize}

The rule i) covers a basic scenario to ensure that all blocks are extended on chain head. The rule ii) could handle conflicting blocks scenario when multiple validators propose blocks in current time slot. The rule iii) ensures the liveness, so that there is at least one block is available for chain extension even if none validator is able to propose a new block.

%---------------------------------------------- Chain Finality Mechanism --------------------------------------
%Reason:1) fork blocks lead to fork branch 2) attacker propose multiple in one slot to generate fork. So we need finality mechanism to resolve fork branch and finalize history of blocks.
\section{Voting-based Chain Finality Mechanism}
\label{sec:BFT} 

Although the block proposal process follows the pre-defined chain extension rules, the fork issue still happens owing to network latency or attacks like "nothing at stake". In addition, through forcing reorganization of the block chain, attacks could change transactions of ledger to launch double-spending attack. Furthermore, the cost of attack may have been lower, since attackers just need to accumulate amount of credit value to be able to propose blocks with high probability, rather than be punished when their violating behaviors are been detected. To prevent above issues, a voting-based chain finality mechanism is introduced to solidify the block chain history and increase cost of attack behaviors by way of properly designing punishment strategies. Our chain finality mechanism borrows idea from Casper \cite{buterin2017casper}, which is a PoS-based finality system overlaying an existing PoW blockchain.

\subsection{Checkpoint Finality Protocol}

%fig2_fig3_checkpoint_tree
\begin{figure} [t]
\begin{center}
\begin{tabular}{c}
\includegraphics[height=6.5 cm]{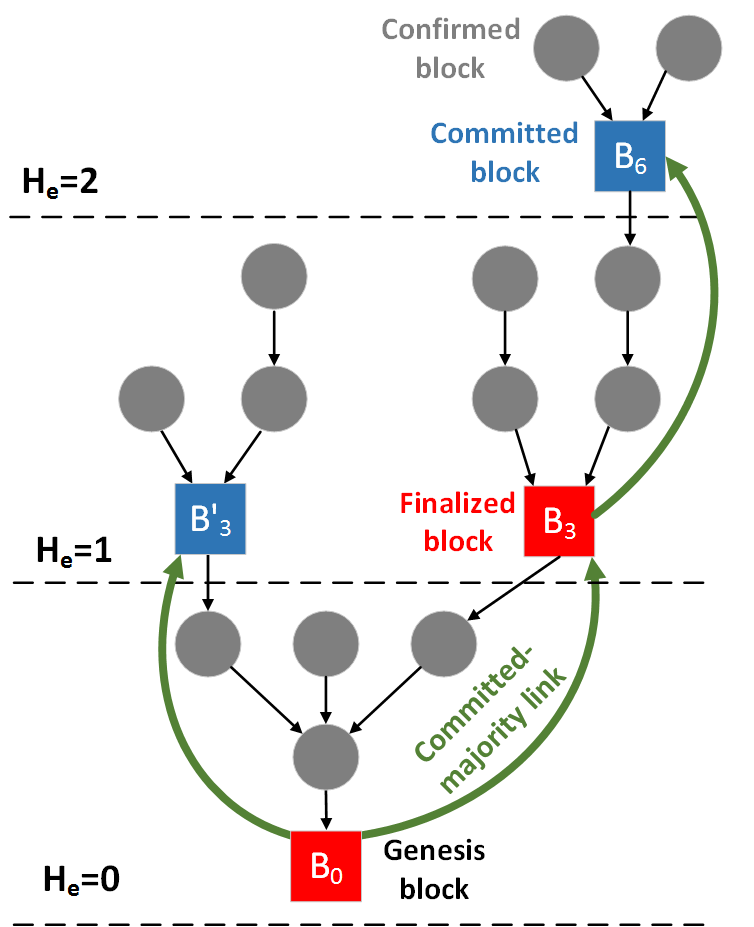}
\end{tabular}
\end{center}
%\caption[example] { \label{fig:checkpoint_tree} a) Chain finality protocol based on checkpoint tree; b) Illustration of chain fork resolving rules.}
\caption[example] { \label{fig:checkpoint_tree} Chain finality protocol based on checkpoint tree}
\vspace{-10pt}
\end{figure}

Under ideal scenarios, the PoC-based block proposal mechanism will propose blocks one after other and extend chain in a linked list, which each parent block has exactly one child block. However, owing to fork issues caused by network latency or deliberate attacks, the block proposal mechanism will inevitably produce multiple conflicting blocks which are children blocks with the same parent block. Therefore, those proposed blocks in fact form an ever-growing \emph{block tree} structure, as shown in Fig. \ref{fig:checkpoint_tree}. The chain finality protocol is mainly to identify a unique chain path on block tree by choosing a single child block from multiple children blocks with common parent block. For efficiency purposes, the chain finality protocol is only executed on those \emph{checkpoint} blocks rather than whole block tree, and committee members vote for hashes of blocks instead of entire block contents. Before describing the voting-based chain finality protocol, several basic definitions are defined as following:
\begin{defn}
Checkpoint - An block $B_i$ whose $height$ in block tree is multiple of epoch size $R$ is specified as checkpoint. If a block is checkpoint, then $B_i[height] \mod R=0$.
\end{defn}

\begin{defn}
Epoch Height - indicates current height of epoch for block $B_i$, which is denoted as $H_e(B_i)=\lfloor \frac{B_i[height]}{R} \rfloor$. 
\end{defn}

\begin{defn}
Vote - A vote for checkpoint block $target$ sent by validator $v_j$ at time $T_{stamp}$ is represented as $Vote_j=(vote\_hash, h_{source}, h_{target},H_e(source),$ $H_e(target), T_{stamp}, pk_j, \sigma_j)$. where the parameters are:

\begin{itemize}
\item $vote\_hash$: is a $\lambda$-bit-length hash string of $Vote_j$, which is represented as $\mathcal{H}(Vote_j)$=$\mathcal{H}(h_{source}, h_{target}, $ $H_e(source), H_e(target), T_{stamp}, pk_j)$, where $\mathcal{H}(\cdot)$ is a pre-defined collision-resistant hash function that output hash string $h \in \{0,1\}^{\lambda}$;
\item $h_{source}$: the hash of any committed block;
\item $h_{target}$: the hash of a checkpoint block which is the a decedent of $source$ block;
\item $H_e(source)$: epoch height of source block $B_s$;
\item $H_e(target)$: epoch height of target block $B_t$;
\item $T_{stamp}$: time stamp the vote is proposed;
\item $pk_j$: public key of validator that sends the vote; and
\item $\sigma_j$: a signature denoted as $Sign_{sk_j}(vote\_hash, h_{source}, h_{target}, H_e(target),$ $ H_e(source), T_{stamp})$ signed by sender $pk_j$.
\end{itemize}
\end{defn}

During time slot $sl_t$ of chain finality process, every validator $v_j$ in current dynasty casts a $vote_j$ message for committing a checkpoint block. Given the voting result, the checkpoint block has following proprieties:
\begin{itemize}
\item \emph{Committed-majority Link}: Given an ordered pair of checkpoint blocks $(B_s, B_t)$ denoted as $B_s \rightarrow B_t$, if more than $2/3$ of validators propose vote for it, then relationship $B_s \rightarrow B_t$ is called a committed-majority link. As Fig. \ref{fig:checkpoint_tree} shown, both $B_0 \rightarrow B_3$ and $B_3 \rightarrow B_6$ are committed-majority link.
\item \emph{Committed Block}: A checkpoint block $B_t$ is called committed block if 1) $B_t$ is genesis block, or 2) there exists a committed-majority link $B_s \rightarrow B_t$ where $B_s$ is a committed block.
\item \emph{Finalized Block}: A checkpoint block $B_s$ is called finalized block if 1) $B_s$ is genesis block, or 2) $B_s$ is a committed block, and there exists a committed-majority link $B_s \rightarrow B_t$, where $B_s$ is parent of $B_t$ in checkpoint tree ($H_e(B_t)=H_e(B_s)+1$).
\end{itemize}

The voting-based chain finality protocol utilizes checkpoint blocks to form a checkpoint tree to finalize chain history. Figure \ref{fig:checkpoint_tree} illustrates a checkpoint tree with a finalized chain along commit-majority links $B_0 \rightarrow B_3 \rightarrow B_6$. The gray nodes represent all confirmed blocks generated by block proposal mechanism, while all checkpoint blocks are shown as squire boxes. Assuming that epoch size $R=3$, all blocks with $(height \mod 3)=0$ are considered as checkpoint blocks, such as $B_0$, $B_3$, $B^{'}_{3}$ and $B_6$. The $B_0$ and $B_3$ are finalized blocks while $B^{'}_{3}$ and $B_6$ are committed blocks. The chain finality workflow is described as follows:

\begin{itemize}
\item \emph{Sending Votes}: during current time slot $sl_t$ for chain finality process, a validator $v_j$ in current dynasty checks if $H_e(head(\mathcal{C}))=H_e(B_c)+1$, where $B_c$ is the last committed block of local chain $\mathcal{C}$. If yes, it prepares $Vote_j$ which votes for a new committed-majority link $B_c \rightarrow head(\mathcal{C})$, and broadcasts signed vote message $Vote_j$ to committee members.  

\item \emph{Counting Votes}: After receiving proposed vote $Vote_j$, each validator $v_i$ ($i \in K$) check if $Vote_j$ is from a valid validator and $\sigma_j$ is correctly signed by sender's $pk_j$. Then, the validator will verify whether received vote violates the chain fork resolving rules defined in \ref{sec:chainforkrule}. If no violation behavior is founded, the validator will updates $vote\_count[h_{source}, h_{target}]$ value by adding 1. Each validator keeps tracking votes count until $sl_t$ expiration.

\item \emph{Finalizing Checkpoint}: A committed-majority link $B_s \rightarrow B_t)$ is only accepted when $vote\_count$ is more than a certain threshold $\mathcal{T} \cdot K$, where $\mathcal{T}$ is a fraction of the committee size $K$. The chain finality process requires that more than $2f+1$ of participants to be honest with $f$ fault nodes in current committee, so that $\mathcal{T} \geq \frac{2f+1}{N}$ where $N=3f+1$. Given $\mathcal{T}=\frac{2}{3}$, if $vote\_count > \frac{2K}{3}$ ($K$ is committee size), the checkpoint block $B_t$ is accepted as a committed block, whereas committed block $B_s$ is changed to a finalized block. %If none of $vote\_count$ is more than $\frac{2K}{3}$ until $sl_t$ expiration, the voting committed-majority link with largest $vote\_count$ will be accepted by validator.
\end{itemize}

Given assumption that time slot $sl_t \geq \Delta$, so that all validators could receive broadcasted votes by end of bounded delay. If no more than $\frac{1}{3}$ of the validators violate the chain fork resolving rule, only a committed block is finalized even if there are conflicting checkpoint blocks in same epoch height. 

\subsection{Chain Fork Resolving Rule}
\label{sec:chainforkrule}

%fig4_fork_rule
\begin{figure} [t]
\begin{center}
\begin{tabular}{c}
\includegraphics[height=6.5 cm]{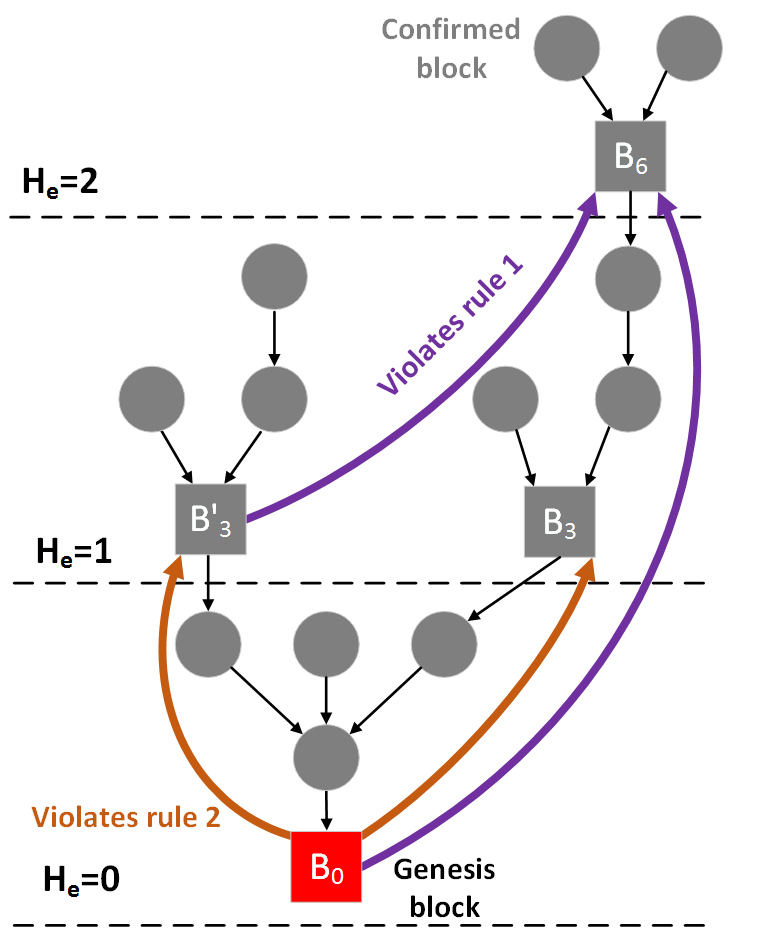}
\end{tabular}
\end{center}
\caption[example] { \label{fig:fork_rule} Illustration of chain fork resolving rules.}
\vspace{-10pt}
\end{figure}

If two checkpoint blocks $B_i$ and $B^{'}_{i}$ have the same epoch height $H_e$ but are in distinct branches, they are \emph{conflicting} checkpoint blocks. In other words, the  checkpoint blocks in checkpoint tree are sibling nodes who are neither ancestors nor descendants of each other. As shown in Fig. \ref{fig:checkpoint_tree}, $B_3$ and $B^{'}_{3}$ are conflicting checkpoint blocks in $H_e=1$. To handle conflicting checkpoint blocks and resolve the fork condition in the chain finality process, a set of chain fork resolving rules are defined as follows:

\begin{ruln}
Given a vote $Vote_j=(vote\_hash, h_{s}, h_{t},$ $H_e(s), H_e(t), T_{stamp}, pk_j, \sigma_j)$, if $H_e(t) \neq H_e(head(\mathcal{C})$ or $H_e(t) \neq H_e(s)+1$, the voter $j$ violates rule and $Vote_j$ will be rejected. This rule ensures that vote for committed-majority $s \rightarrow t$ is accepted only if $t$ is the same epoch height as local chain head and $s$ is parent of $t$ in checkpoint tree.  
\end{ruln}

\begin{ruln}
Given two votes $(vote\_hash, h_{s1}, h_{t1},$ $H_e(s1), H_e(t1), T_{stamp}, pk_j, \sigma_j)$ and $(vote\_hash, h_{s2}, h_{t2},$ $H_e(s2), H_e(t2), T_{stamp}, pk_j, \sigma_j)$ from the same voter $j$, if $t1 \neq t2$ and $H_e(t1) = H_e(t2)$, the voter $j$ violates rule and all votes from voter will be rejected. This rule prevents one voter from voting for distinct checkpoint blocks with the same epoch height.  
\end{ruln}

Figure \ref{fig:fork_rule} demonstrates scenarios that how voter violates above defined fork rules. A validator who voted for distinct target blocks $B_3$ and $B^{'}_{3}$ in the same epoch height violates the rule 2. Since $B^{'}_{3}$ is not ancestor of $B_6$, any sender voting for link $B^{'}_{3} \rightarrow B_6$ violates rule 1. Similarly, voting for link $B_0 \rightarrow B_6$ is also considered as violation of rule 1 owing to condition $H_e(B_6) \neq H_e(B_0)+1$.

\subsection{Incentives}
%This section provides a brief discussion on incentive mechanism of protocol which covers both reward and punishment strategy. 
In order to encourage participants to contribute honest protocol operations to distributed ledger, a reward can be given to the validators that are proposing, verifying blocks and voting for finalized checkpoint blocks. The rewards in our setting are fees and credit value. At end of dynasty epoch, all fees of transactions included in a sequence of blocks in current committee are collected as a rewarding fees pool, then it will be distributed to all honest validators who have no record of violating behaviors. In addition, each honest validator $v_i$ will also be added 1 credit value to his credit stake $c_i$. The higher credit stake $c$, the more benefit that participant can earn by being selected as committee member and committing honest protocol actions.   

Given violations of chain fork resolving rules, a punishment strategy is also introduced to discourage dishonest behaviors. To become a committee member in committee selection process, a participant $v_i$ must deposit a fix number of fees to his special stake $sc_i$, called \emph{security stake}. If any dishonest behaviors of validator $v_i$ are detected, such as violations of chain fork resolving rules, his balance of $sc_i$ will be slashed as punishment. Furthermore, validator $v_i$ will also lose 1 credit value from his credit stake $c_i$. Such punishment strategy not only increase economical cost of dishonest behavior by slashing his security stake, but also decrease the capability of attackers by lowing their credit stakes.

%In order to encourage participants to a contribute honest protocol operations to distributed ledger, a reward can be given to the validators that are proposing, verifying blocks and voting for finalized checkpoint blocks. At end of dynasty epoch, all honest validators who have no record of violating behaviors will received 1 credit value increase to their credit stake $c$. Given violations of chain fork resolving rules, a punishment strategy is also introduced to discourage dishonest behaviors. If any dishonest behaviors of validator $v_i$ are detected, 1 credit value will be deducted from his credit stake $c_i$. Such punishment strategy could decrease the capability of attacker by lowing his credit stake. The lower credit stake $c$ makes attackers have the lower probability to be selected as committee members and participate in the block proposal and chain finality process.       

\section{Implementation and Evaluation}
\label{sec:experiment} 
% ------------------------------ implementation and experimental setup -------------------------------------
To verify the proposed solution, a concept-proof prototype of microchain is implemented in Python, consisting of approximate 3000 lines of code. We use Flask \cite{flask}, which provide a micro-framework for Python application, to develop networking and web service functions. All security functions are developed by using standard python lib: cryptography \cite{pyca}. The key generation and signature are implemented over RSA and hash function uses SHA-256. we use SQLite\cite{sqlite}, which is an lightweight and embedded SQL database engine, to manage data such as node, block and vote information.

The prototype is deployed on a physical network environment including multiple nodes. Table \ref{tab:testbed} describes devices used for the experimental setup. Five validators are deployed on desktop while other validators run on sixteen distributed Raspberry Pis to simulate IoTs. Each validator is only deployed on one host machine.

\begin{table}[ht]
\caption{Configuration of Experimental Nodes.} 
\label{tab:testbed}
\begin{center}       
\begin{tabular}{|l|p{2.5cm}|p{3.5cm}|} %% this creates two columns
%% |l|l| to left justify each column entry
%% |c|c| to center each column entry
%% use of \rule[]{}{} below opens up each row
\hline
\rule[-1ex]{0pt}{3.5ex} \textbf{Device} & Dell Optiplex 760 & Raspberry Pi 3 Model B+ \\
\hline
\rule[-1ex]{0pt}{3.5ex} \textbf{CPU} & 3 GHz Intel Core TM (2 cores) & Broadcom ARM Cortex A53 (ARMv8) , 1.4GHz \\
\hline
\rule[-1ex]{0pt}{3.5ex} \textbf{Memory} & 4GB DDR3 & 1GB SDRAM \\
\hline
\rule[-1ex]{0pt}{3.5ex} \textbf{Storage} & 250G HHD & 32GB (microSD card) \\
\hline
\rule[-1ex]{0pt}{3.5ex} \textbf{OS} & Ubuntu 16.04 & Raspbian GNU/Linux (Jessie) \\
\hline
\end{tabular}
\end{center}
\end{table}

%------------------------------ Performance Evaluation -------------------------------------
\subsection{Network Latency}
% fig4_network_latency
\begin{figure} [t]
\begin{center}
\begin{tabular}{c}
\includegraphics[height=4.8 cm]{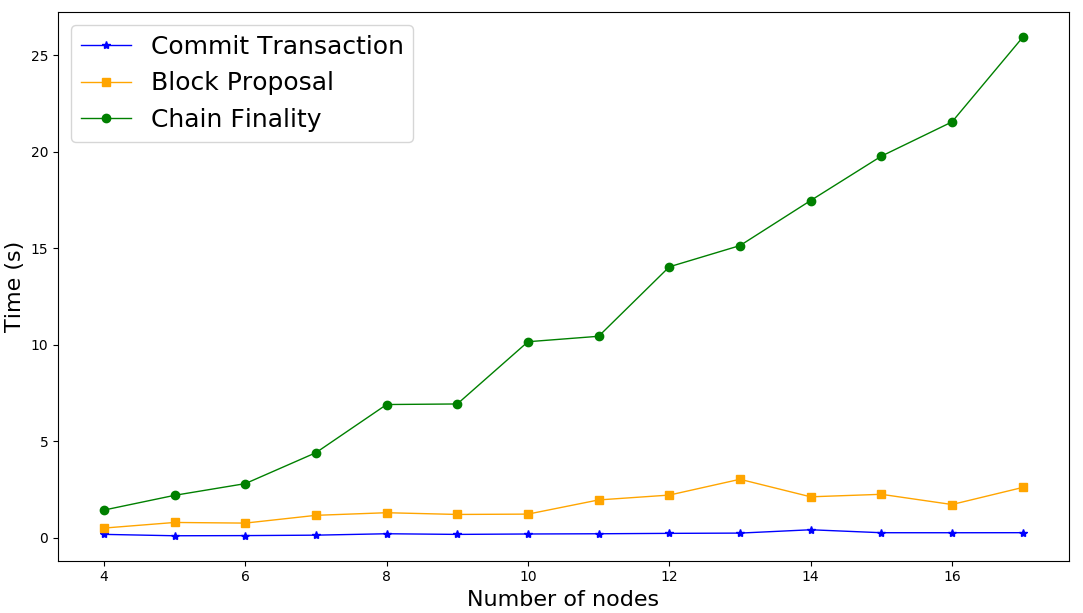}
\end{tabular}
\end{center}
\caption[example] { \label{fig:network_latency} Latency for one round of microchain with different node size.}
\vspace{-10pt}
\end{figure}

In this experiment, all validators are running on Raspberry Pis to evaluate network latency of executing microchain on IoT devices. Figure \ref{fig:network_latency} plots the time that takes for microchain to complete a entire round of final-committee consensus given the number of validators varying from 4 to 16. The block size is 128KB in test to focus on network performance by reducing block size influence. 

The whole process of one round is divided into three parts, which is demonstrated as three lines. The latency of commit transaction $\mathcal{T}_{ct}$, at the bottom of Fig. \ref{fig:network_latency}, is the time for all nodes of dynasty to accept a broadcasted transaction. Since the communication complexity of broadcasting transaction is $\mathcal{O}(K)$. The latency of committing transaction is linear scale to committee size $K$, and it varies from 162 ms to 246 ms. 

The line in the middle of graph indicates latency of block proposal $\mathcal{T}_{bp}$, which measure how long the proposed blocks could be arrived and verified among validators. Since the block proposal algorithm is proportion to credit distributuion $\mathcal{D}$ whth expectation $E(\mathcal{D})$, the latency of block proposal is scale to communication complexity $\mathcal{O}(\frac{K^2}{E(\mathcal{D})})$. Given uniform distribution $\mathcal{D}$ in our test with $E(\mathcal{D})=K$, so that the middle line is almost linear scale to the $\mathcal{O}(K)$ with vary from 0.5 s to 1.7 s. 

Finally, the latency of chain finality, at the top of the graph, is the time it takes voting process for finalize checkpoint block to complete among all nodes. Since voting-based chain finality process has the communication complexity $\mathcal{O}(K^2)$, the top line indicates the latency of chain finality $\mathcal{T}_{cf}$, which is greatly influenced by the number of nodes. Given 16 validators in committee, the latency could be 21.5 s, while the scenario with four nodes only introduces 1.4 s latency.

The results show that microchain could perform one round of final-committee consensus no more than half minute $(\mathcal{T}_{ct} + \mathcal{T}_{bp}+\mathcal{T}_{cf})$ given committee size $m=16$. Although chain finality process introduces larger latency as increasing committee size, properly configuring epoch size $R$ could mitigate the influence. The block confirmation time denotes as $T_{bc}=\frac{\mathcal{T}_{cf}+\mathcal{T}_{bp}\cdot R}{R}$, given $R=10$ so that chain finality is triggered every 10 block height increases, $T_{bc}=(21.5 + 2.5 \times 10 )/10 \approx 4.6 $ s. Compared with block generation time of Ethereum blockchain with 9.35 s in private network \cite{xu2018smartcac} and about 15 s in public network, the proposed microchain mechanism demonstrates lower block confirmation time. Since the block confirmation time of network is normalized to committee size $K$, properly choose committee size and designing committee selection algorithm could improve scalability of microchain in IoT applications.

\subsection{Throughput Evaluation}

% fig5_blocksize_latency
\begin{figure} [b]
\vspace{-10pt}
\begin{center}
\begin{tabular}{c}
\includegraphics[height=4.6 cm]{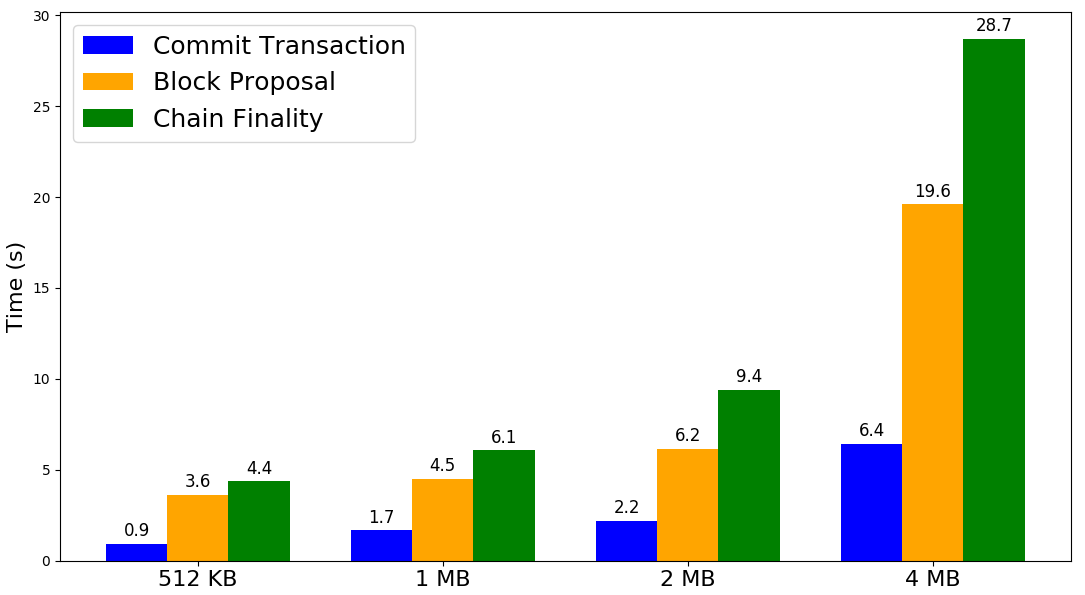}
\end{tabular}
\end{center}
\caption[example] { \label{fig:blocksize_latency} Latency for one round of microchain with different block size.}
\end{figure}

In the following set of experiments, we deploy five validators in committee to focus on throughput and cost evaluation by limiting committee size influence, which are all Raspberry Pi devices. Figure \ref{fig:blocksize_latency} plots the time that takes for microchain to complete a entire round of final-committee consensus with a varying block size between 512K and 4M. The results show that latency of running consensus algorithm is scale to the block size in microchain. Considering the scenario with lowest latency, microchain finalizes a 512 KBytes block in about 8.9 s, which means it could finalize 202 Mbytes of transaction data per hour. For comparison, Bitcoin achieves throughput of processing 6 MBytes transaction data per hour by committing a 1 Mbytes block per 10 minutes.

The throughput could be specified as $Th=\frac{\mathcal{T}_{ct} + \mathcal{T}_{bp}+\mathcal{T}_{cf}}{Block\_size} \times 3600$ (M/h), where M/h means Mbytes per hour. Then we can get throughput with varying block size: $Th_{512K}$=202 (M/h), $Th_{1M}$=293 (M/h), $Th_{2M}$=405 (M/h), and $Th_{4M}$=263 (M/h). The above throughput results with varying block size suggests that increasing the block size allows committing more transaction data, and therefore reach a throughput which maximizes the system capability. In our test, running microchain with 2M block size outputs highest throughput 405 (M/h). As block size increases, however, microchain achieves higher throughput at the cost of increasd latency, and throughput is constrained by network and system capability.

\subsection{Performance of Running Microchain}

\begin{figure} [t]
\begin{center}
\begin{tabular}{c}
\includegraphics[height=4.6 cm]{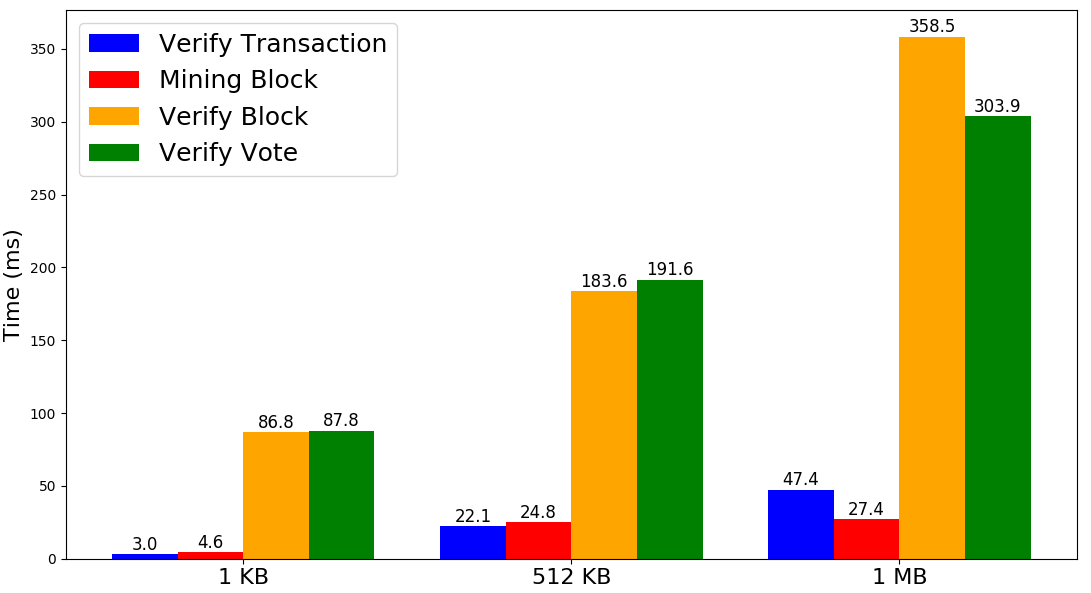}
\end{tabular}
\end{center}
\caption[example] { \label{fig:blocksize_performance} Performance of running microchain with different block size.}
\vspace{-10pt}
\end{figure}

To evaluate the cost of performing microchain on host machine, running time for key operations are calculated given different block size and hardware platforms.
%larger block size, higher cost of running time.
Figure \ref{fig:blocksize_performance} demonstrates the computational overhead of key tasks for completing a entire round of final-committee consensus on Raspberry Pi with varying block size. The results show that computation overhead increases as bloc size grows. Compared with verify transaction and mining block operations, the verify block and verify vote operations introduces major overhead to host machine. Since verify transaction and mining block process rely on database querying to validate chain data, therefore they require more computing resource on data processing and are sensitive to block size growth. 

\begin{figure} [t]
\begin{center}
\begin{tabular}{c}
\includegraphics[height=4.6 cm]{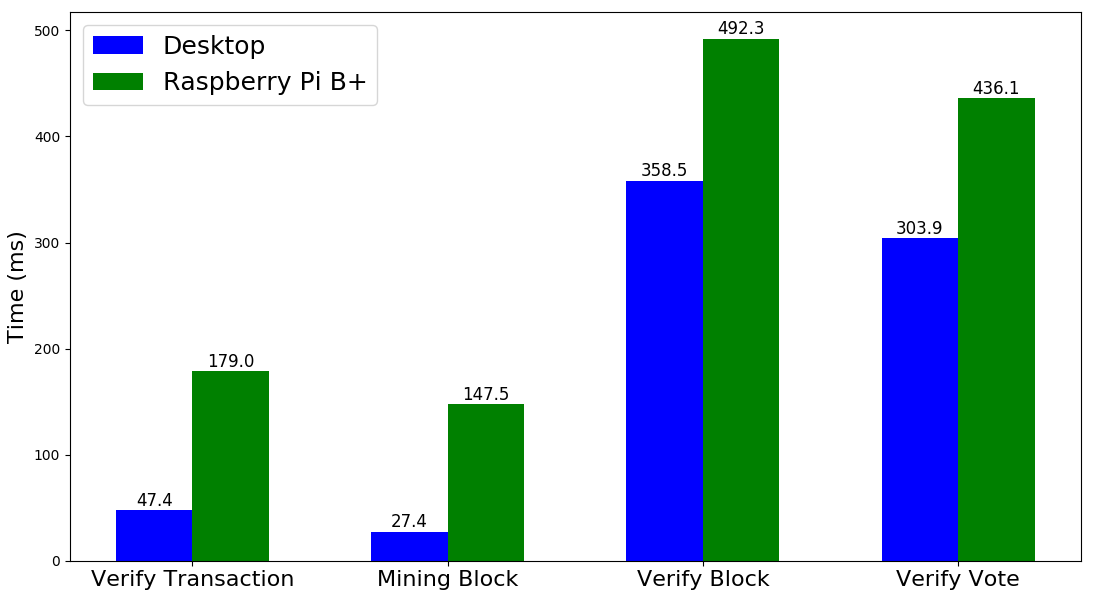}
\end{tabular}
\end{center}
\caption[example] { \label{fig:platform_performance} Performance of running microchain with different platform.}
\vspace{-10pt}
\end{figure}

Figure \ref{fig:platform_performance} show the cost of running microchain with 1M block size on both desktop and Raspberry Pi. All key tasks for completing a entire round of final-committee consensus introduces higher computational cost on Raspberry Pi B+ owning to constrained resource. The CPU usage for running microchain is modest on both platform. It is about 15\% on Raspberry Pi B+ and 10\% on desktop, and most of CPU consumption is used for verifying signature, calculating hash value and access to the database. Compared with PoW-based consensus protocol that occupies almost all CPU usage, proposed PoC-VCF based final-committee consensus in microchain only introduced lighter overhead on the host machine.

%erify block need query database, so they need more time than verify mining block.

% fig6_blocksize_performance

% fig7_performance_platform

\subsection{Security Discussions}

Since the adversary is subject to the usual cryptographic hardness assumptions, he/she is aware of neither the private keys of the honest nodes nor the input string $x$ to the VRF function. Therefore, the unpredictability property of the VRF-based randomness string generation allows members of committee be completely random. In addition, the chain finality requires $n \geq 3f+1$ to make agreements on checkpoints. Therefore, the adversary has at most $m = 1/4$ chance per round to control the checkpoint voting process. As a result, the probability that an adversary controls $n$ consecutive checkpoint is upper-bounded by $P[X \geq n]=\frac{1}{4^n}<10^{-\lambda}$. For $\lambda = 6$, the adversary will control at most 10 consecutive chain finality runs. 

The security proprieties of micorchain could mitigate the following attacks:
\begin{itemize}
\item \emph{Double spending attacks}: In a double spending attack, the adversary attempts to revert a transaction which is confirmed in the blockchain network. The chain finality ensures persistence, so that once transaction is finalized in checkpoint block, all other honest nodes will work on finalized chain and disregard the double spending transaction. 

\item \emph{Transaction denial attacks}: The adversary could start a transaction denial attack by preventing a certain transaction from becoming confirmed. However, such a attack is not feasible owing to the liveness provided by microchain. Once a transaction is successfully enclosed in a proposed block for a sufficient number of time slots, it will be eventually finalized by the chain finality process.

\item \emph{Long-range attacks}: In a long-range attacks, the adversary wishes to double spend at a later point of time by locally computing a longer valid chain that starts from genesis block and presenting such longer alternative chain to revert his previous confirmed transaction. Since the committee selection is unpredictable, and any blocks that are not proposed by committee members will be rejected. Therefore, it's difficult for the adversary to control the committee selection, and his local chain cannot be accept by committee member. Furthermore, the chain finality ensures that valid blocks are only extended alone longest finalized chain, so that the adversary's alternative chain is invalid and will be rejected. 

\item \emph{Selfish-mining}: In a selfish-mining attack, the attacker tries to withhold blocks and release them strategically to reduce chain growth and increase the relative ratio of his proposed blocks. However, our reward and punishment mechanism make such attack become unprofitable. Only honest validators without violating behavior will receive rewards proportionally to their contributions. The rational validators attempt to behave honestly and make contribution to increase credit stake.
\end{itemize}

% merged as one: fig_experiment
%\begin{figure} [t]
%\begin{center}
%\begin{tabular}{c}
%5\includegraphics[height=5.4 cm]{fig_experiments.png}
%\end{tabular}
%\end{center}
%\caption[example] { \label{fig:fig_experiment} Experimental results.}
%\vspace{-10pt}
%\end{figure}
%Figure \ref{fig:fig_experiment}

\section{Conclusions}
\label{sec:conclusion}  % \label{} allows reference to this section
This work proposes microchain, a hybrid blockchain network that provides scalable and lightweight distribute ledger for IoT systems. Through performing PoC-VCF consensus protocol in a small group of validators, called final-committee, microchain could effectively reduce communication complexity and improve performance, i.e. in terms of confirmation time and throughput. The VRF-based cryptographic sortition and unbiased randomness generation allow committee selection be unpredictable, therefore it increases scalablity of microchain. The experimental results demonstrate that microchain achieves higher throughput than PoW-based blockchain solutions, and it introduces lower network latency and modest computational overhead on IoT devices like Raspberry Pi. The microchain offers a prospective distributed ledger solution to IoT application scenarios in term of security, scalability and lightweight. 

There remains a number of open issues in designing microchain.  
%Although incentives are considered in microchain design, more design and analysis are necessary to encourage users to participate network and prevent dishonest behaviors. Another challenge is redesign chain structure to address ever-growing chain data which has great impact on computation and storage capability of IoT devices.
Although committee selection could improve scalability of microchain, more investigation and test are needed to evaluate how committee selection algorithm scale to network size. Another challenge is redesigning chain structure to address the ever-growing chain data size, which has a great impact on computation and storage capability of IoT devices.

%\appendices

%\section{}

%\ifCLASSOPTIONcompsoc
%\else
%\fi


\ifCLASSOPTIONcaptionsoff
  \newpage
\fi



%\bibliographystyle{IEEEtran}
%\begin{thebibliography}{1}


%\end{thebibliography}

\bibliographystyle{IEEEtranS} % makes bibtex use IEEEtran.bst
\bibliography{report} % bibliography data in report.bib


%\begin{IEEEbiographynophoto}{Ronghua Xu}
\begin{IEEEbiography}[{\includegraphics[width=1in,height=1.25in,clip,keepaspectratio]{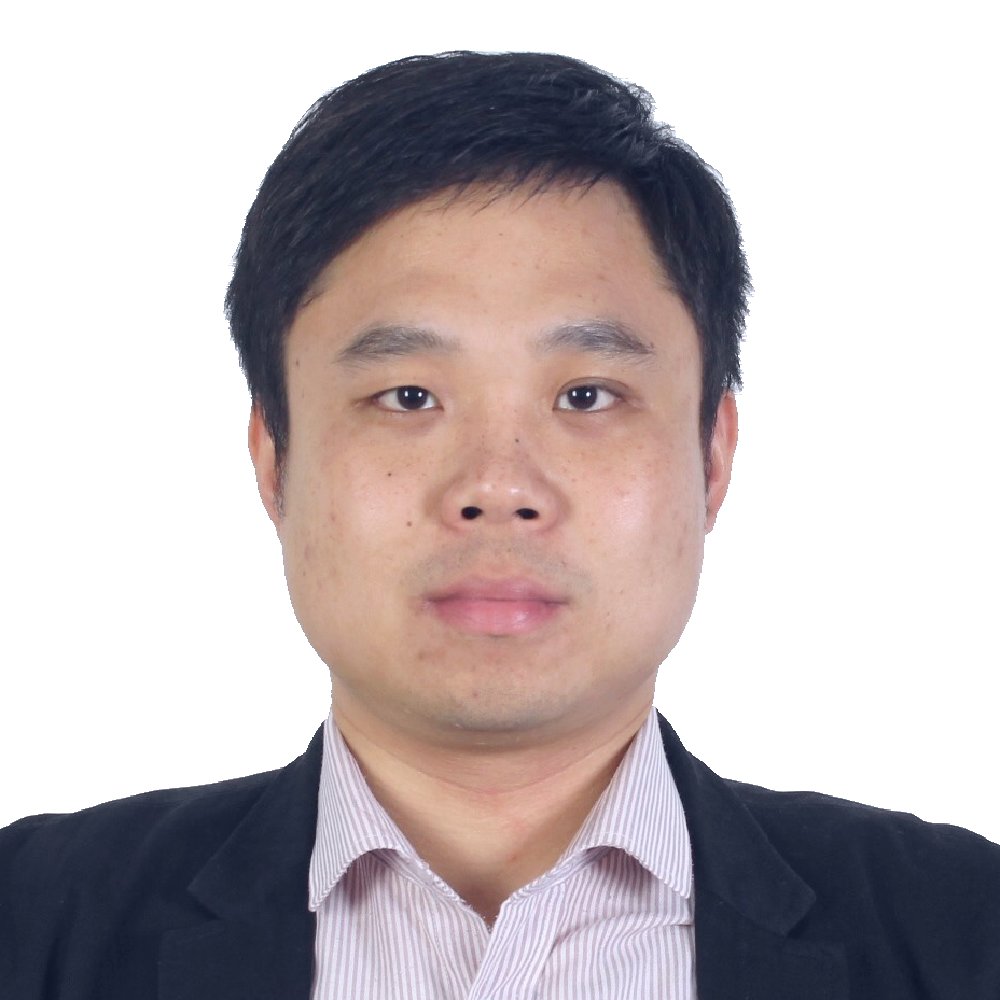}}]{Ronghua Xu}
is a Ph.D student of Electrical and Computer Engineering at the Binghamton University - State University of New York (SUNY). He earned his B.S. on Mechanical Engineering from Nanjing University of Science \& Technology, China in 2007, and received his M.S. degree on Mechanical and Electrical Engineering from Nanjing University of Aeronautics \& Astronautics in 2010.His research interest lies in Machine Learning; Blockchain, Algorithm Design; Cloud/Fog/Edge Computing Paradigm. His email address is rxu22g@binghamton.edu.
\end{IEEEbiography}

\begin{IEEEbiography}[{\includegraphics[width=1in,height=1.25in,clip,keepaspectratio]{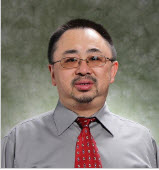}}]{Yu Chen}
is an Associate Professor at the Binghamton University-SUNY. He received the Ph.D. in Electrical Engineering from the University of Southern California (USC) in 2006. His research interest lies in Trust, Security and Privacy in Computer Networks, focusing on Edge-Fog-Cloud Computing, Internet of Things (IoT), and their applications in smart and connected environments. His publications include over 150 papers in scholarly journals, conference proceedings, and books. His email address is ychen@binghamton.edu.
\end{IEEEbiography}

%\newpage

\begin{IEEEbiography}[{\includegraphics[width=1in,height=1.25in,clip,keepaspectratio]{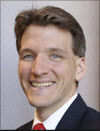}}]{Erik Blasch}
is a Program Officer at the Air Force Office of Scientific Research (AFOSR). He received his BS from MIT, six MS degrees, and PhD from Wright State University. He has compiled 800+ papers, 5 books, 27 patents, and 30 tutorials. He is a recipient of the Military Sensing Society Mignogna data fusion award, past president of the International Society of Information Fusion, AIAA Associate Fellow, SPIE Fellow, and IEEE Fellow. His email address is erik.blasch.1@us.af.mil.
\end{IEEEbiography}

\begin{IEEEbiography}[{\includegraphics[width=1in,height=1.25in,clip,keepaspectratio]{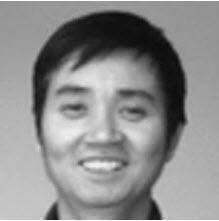}}]{Genshe Chen}
is the CTO of Intelligent Fusion Technology, Inc., at
Germantown, Maryland. He received his BS and MS degrees in electrical
engineering and his PhD in aerospace engineering, in 1989,
1991, and 1994, respectively, all from Northwestern Polytechnical
University, Xian, China. He was a postdoctoral fellow at Beijing University
of Aeronautics and Astronautics and Wright State University
from 1994 to 1997. He was with Intelligent Automation, Inc., Rockville,
Maryland, from 2004 to 2007.
\end{IEEEbiography}

\vfill


\end{document}